\documentclass[final]{IEEEtran} 

\usepackage{amssymb,amsmath,bm}

\usepackage{graphicx,color}

\usepackage{cite}
\usepackage{url}

\newcommand\mymatrix[1]{\bm{\mathrm{#1}}}

\newlength\imagewidth
\newlength\figwidth
\newlength\figsep
\begin{document}

\title{Cryptanalysis of an Encryption Scheme Based on Blind Source Separation%
\thanks{This research was partially supported by The Hong Kong Polytechnic
University's Postdoctoral Fellowships Program under grant no.
G-YX63. The work of K.-T. Lo was supported by the Research Grants
Council of the Hong Kong SAR Government under Project Number 523206
(PolyU 5232/06E).}}

\author{Shujun Li\thanks{Shujun Li and Kwok-Tung Lo are with the Department
of Electronic and Information Engineering, The Hong Kong Polytechnic
University, Hung Hom, Kowloon, Hong Kong SAR, P.~R. China.},
Chengqing~Li\thanks{Chengqing Li and Guanrong Chen are with the
Department of Electronic Engineering, City University of Hong Kong,
Kowloon Toon, Hong Kong SAR, P.~R. China.},
Kwok-Tung~Lo,~\IEEEmembership{Member, IEEE} and
Guanrong~Chen,~\IEEEmembership{Fellow, IEEE}\thanks{The
corresponding author is Shujun Li. Contact him via his person web
site: http://www.hooklee.com.}}

\maketitle

\begin{abstract}
Recently Lin et al. proposed a method of using the underdetermined
BSS (blind source separation) problem to realize image and speech
encryption. In this paper, we give a cryptanalysis of this BSS-based
encryption and point out that it is not secure against
known/chosen-plaintext attack and chosen-ciphertext attack. In
addition, there exist some other security defects: low sensitivity
to part of the key and the plaintext, a ciphertext-only differential
attack, divide-and-conquer (DAC) attack on part of the key. We also
discuss the role of BSS in Lin et al.'s efforts towards
cryptographically secure ciphers.
\end{abstract}

\section{Introduction}

With the rapid development of multimedia and networking
technologies, the security of multimedia data becomes more and more
important in many real applications. To fulfill such an increasing
demand, during past decades many encryption schemes have been
proposed to protect multimedia data, including speech, images and
videos \cite{Beker:SecureSpeech:Book1985,
Kumar:SecureCryptology:Book1997, Nichols:SpeechCryptology:Book2002,
Furht:ImageVideoEncryption:Handbook2004,
Li:ChaosImageVideoEncryption:Handbook2004,
Uhl:ImageVideoEncryption:Book2005,
Furht:MultimediaSecurity:Book2005, Zeng:MultimediaSecurity:Book2006,
Javidi:OpticalSecurity2005}.

According to the nature of protected data, multimedia encryption
schemes can be classified into two basic types: analog and digital.
Most early schemes were designed to encrypt analog data in various
ways: element permuting, signal masking, frequency shuffling, etc.,
all of which may be exerted in time domain or transform domain or
both. However, due to the simplicity of the encryption procedures,
almost all analog encryption schemes are not sufficiently secure
against cryptographical attacks, especially those modern attacks
such as known/chosen-plaintext and chosen-ciphertext attacks
\cite{Kumar:SecureCryptology:Book1997,
Nichols:SpeechCryptology:Book2002, BreakingNagravision:1999,
Li:AttackingPOMEA2004}. As a comparison, in digital encryption
schemes, one can employ any cryptographically strong cipher, such as
DES \cite{Schneier:AppliedCryptography96} or AES
\cite{NIST:AES2001}, to achieve a higher level of security. Besides,
to achieve a higher efficiency of encryption and some special
demands of multimedia encryption (such as format-compliance
\cite{Zeng:VideoScrambling:IEEETCASVT2002} and perceptual encryption
\cite{Li:PerceptualEncryption:2005}), many specific multimedia
encryption schemes have also been developed
\cite{Furht:ImageVideoEncryption:Handbook2004,
Li:ChaosImageVideoEncryption:Handbook2004,
Uhl:ImageVideoEncryption:Book2005}. Recent cryptanalysis work
\cite{BreakingSFCVideoEncryption:EuroCrypt89,
Jan-Tseng:SCAN:IPL1996, Qiao:IsRandomOrderSecure:ISCE97,
Uehara:ChosenDCTAttack:IEEEPCM2000, Yu-Chang:SCAN:PRL2002,
Youssef:BreakingFEA-M:IEEETCE2003, Li:AttackingFEAM:JSS2006,
Li-Zheng:CKBA:ISCAS2002, Li-Zheng:BRIE:ICIP2002,
LiLi:AttackingCNN2004, Li:EURASIP-JASP2005, LiLiLiChen:ISCAS2005,
Li:JSS2006, Li:AttackingRCES2004, ShujunLi:AttackISWBE2006} has
shown that some multimedia encryption schemes are insecure against
various cryptographical attacks.

Recently Lin et al. suggested employing blind source separation
(BSS) for the purpose of image and speech encryption
\cite{Lin:BSS_IE:IEE_EL2002, Lin:BSS_IE:ICNNSP2003,
Lin:BSS_IE:CASSET2004, Lin:BSS_SIE:ISNN2005, Lin:BSS_IE:ISNN2006,
Lin:BSS_SE:ICCCAS2004, Lin:BSS_SE:IEEETCASI2006}. The basic idea is
to mix multiple plaintexts (or multiple segments of the same
plaintext) with a number of secret key signals, in the hope that an
attacker has to solve a hard mathematical problem -- the
underdetermined BSS problem. In Sec.~VII of
\cite{Lin:BSS_SE:IEEETCASI2006}, Lin et al. claimed that this
BSS-based cipher ``is immune from the attacks such as the
ciphertext-only attack, the known-plaintext, and the
chosen-plaintext attack", ``as long as the intractability of the
underdetermined BSS problem is guaranteed by the mixing matrix for
encryption".

This paper re-evaluates the security of the BSS-based encryption
scheme and points out that it is actually insecure against
known/chosen-plaintext attack and chosen-ciphertext attack. In
addition, some other security defects are also found under the
ciphertext-only attacking scenario, including the low sensitivity to
the mixing matrix (part of the secret key) and the plaintext, and a
differential attack that works well when the matrix size is small.
Based on the cryptanalytic findings, we also discuss the role of BSS
in Lin et al.'s efforts towards cryptographically secure ciphers.

The rest of this paper is organized as follows. In next section we
give a brief introduction to the BSS-based encryption scheme.
Section~\ref{section:Cryptanalysis} is the main body of this paper
and focuses on the cryptanalysis of the BSS-based encryption scheme.
Then, the role of BSS in cryptography is discussed in
Sec.~\ref{section:Discussion}. Finally the last section concludes
this paper.

\section{BSS-Based Encryption}

Blind source separation is a technique that tries to recover a set
of unobserved sources or signals from observed mixtures
\cite{Cardoso:BSS:IEEEProc1998}. Given $N$ unobserved signals
$\mymatrix{s}_1,\cdots,\mymatrix{s}_N$ and a mixing matrix
$\mymatrix{A}$ of size $N\times M$, the BSS problem is to recover
$\mymatrix{s}_1,\cdots,\mymatrix{s}_N$ from $M$ observed signals
$\mymatrix{x}_1,\cdots,\mymatrix{x}_M$, where
\begin{equation}
[\mymatrix{x}_1,\cdots,\mymatrix{x}_M]^T=\mymatrix{A}[\mymatrix{s}_1,\cdots,\mymatrix{s}_N]^T.
\end{equation}
When $M\geq N$, the blind source separation is possible when
$\mymatrix{A}$ satisfies some requirements. However, when $M<N$,
this is generally impossible (whatever $\mymatrix{A}$ is), thus
leading to the underdetermined BSS problem.

In \cite{Lin:BSS_IE:IEE_EL2002, Lin:BSS_IE:ICNNSP2003,
Lin:BSS_IE:CASSET2004, Lin:BSS_SE:ICCCAS2004, Lin:BSS_SIE:ISNN2005,
Lin:BSS_IE:ISNN2006, Lin:BSS_SE:IEEETCASI2006}, Lin et al.
introduced a number of secret key signals to make the determination
of the plaintext signals become an underdetermined BSS problem in
the case that the key signals are unknown. Given $P$ input
plain-signals $s_1(t),\cdots,s_P(t)$ and $Q$ key signals
$k_1(t),\cdots,k_Q(t)$, the encryption procedure is described as
follows\footnote{To achieve a clearer description of the BSS-based
encryption scheme, in this paper we use some notations different
from those in Lin et al.'s original papers. For example, in
\cite{Lin:BSS_SE:IEEETCASI2006}, the $i$-th key signal is denoted by
$s_{ni}(t)$, while in this paper we use $k_i(t)$ to emphasize the
fact that it is a \textbf{key} signal.}:
\begin{equation}\label{equation:encryption0}
\mymatrix{x}(t)=[x_1(t),\cdots,x_P(t)]^T=\mymatrix{A}\mymatrix{s}_k(t),
\end{equation}
where $\mymatrix{x}(t)$ denote $P$ cipher-signals,
$\mymatrix{s}_k(t)=[s_1(t),\cdots,s_P(t),k_1(t),\cdots,k_Q(t)]^T$,
and $\mymatrix{A}$ is a $P\times(P+Q)$ mixing matrix whose elements
are within in $[-1,1]$. Assume that
$\mymatrix{A}=[\mymatrix{A}_s,\mymatrix{A}_k]$, where
$\mymatrix{A}_s$ is a $P\times P$ matrix and $\mymatrix{A}_k$ is a
$P\times Q$ matrix. Then, the encryption procedure can be
represented in an equivalent form:
\begin{equation}\label{equation:encryption}
\mymatrix{x}(t)=\mymatrix{A}_s\mymatrix{s}(t)+\mymatrix{A}_k\mymatrix{k}(t),
\end{equation}
where $\mymatrix{s}(t)=[s_1(t),\cdots,s_P(t)]^T$ and
$\mymatrix{k}(t)=[k_1(t),\cdots,k_Q(t)]^T$. Thus, as long as
$\mymatrix{A}_s$ is an invertible matrix, one can decrypt
$\mymatrix{s}(t)$ as follows\footnote{In Lin et al.'s papers, it is
said that the decryption procedure was achieved via BSS. However,
from the cryptographical point of view, it is more convenient to
denote the decryption procedure by
Eq.~(\ref{equation:decryption}).}:
\begin{equation}\label{equation:decryption}
\mymatrix{s}(t)=\mymatrix{A}_s^{-1}\left(\mymatrix{x}(t)-\mymatrix{A}_k\mymatrix{k}(t)\right).
\end{equation}

Different values of $Q$ was used in Lin et al.'s papers: $Q=1$ in
\cite{Lin:BSS_IE:IEE_EL2002} and $Q=P$ in
\cite{Lin:BSS_IE:ICNNSP2003, Lin:BSS_IE:CASSET2004,
Lin:BSS_SE:ICCCAS2004, Lin:BSS_SIE:ISNN2005, Lin:BSS_IE:ISNN2006,
Lin:BSS_SE:IEEETCASI2006}. When $Q=P$, Lin et al. further set
$\mymatrix{A}_s=\mymatrix{B}$ and
$\mymatrix{A}_k=\beta\mymatrix{B}$, where $\beta\geq 10$ for image
encryption and $\beta\geq 1$ for speech encryption. In this case,
the encryption procedure becomes
\begin{equation}
\mymatrix{x}(t)=\mymatrix{B}\left(\mymatrix{s}(t)+\beta\mymatrix{k}(t)\right),
\end{equation}
and the decryption procedure becomes
\begin{equation}
\mymatrix{s}(t)=\mymatrix{B}^{-1}\mymatrix{x}(t)-\beta\mymatrix{k}(t).
\end{equation}

Observing Eq.~(\ref{equation:encryption}), one can see that the
encryption procedure contains two steps:
\begin{itemize}
\item
\textit{Step 1}:
$\mymatrix{x}^{(1)}(t)=\mymatrix{A}_s\mymatrix{s}(t)$;

\item
\textit{Step 2}:
$\mymatrix{x}(t)=\mymatrix{x}^{(1)}(t)+\mymatrix{A}_k\mymatrix{k}(t)$.
\end{itemize}
The first step corresponds to a substitution (block) cipher, and the
second step corresponds to a additive stream cipher. From another
point of view, the two steps are exchanged as follows:
\begin{itemize}
\item
\textit{Step 1}:
$\mymatrix{x}^{(1)}(t)=\mymatrix{s}(t)+\mymatrix{A}_s^{-1}\mymatrix{A}_k\mymatrix{k}(t)$;

\item
\textit{Step 2}:
$\mymatrix{x}(t)=\mymatrix{A}_s\mymatrix{x}^{(1)}(t)$.
\end{itemize}
In any case, the BSS-based encryption scheme is always a product
cipher composed by a simple block cipher and an additive stream
cipher. In next section, we will show that the two sub-ciphers can
be separately broken by known/chosen-plaintext attack and
chosen-ciphertext attack.

In the BSS-based encryption scheme, the key signals
$k_1(t),\cdots,k_Q(t)$ are as long as the plain-signals and have to
be generated by a pseudo-random number generator (PRNG) with a
secret seed $\mathrm{I}_0$, which serves as the secret key. In Lin
et al.'s papers, it was not explicitly mentioned whether or not the
mixing matrix should be used as part of the secret key. However, if
the attacker knows $\mymatrix{A}$, the product cipher degrades to be
a stream cipher. Considering
$\mymatrix{x}^*(t)=\mymatrix{A}_s^{-1}\mymatrix{x}(t)$ as the
equivalent cipher-signal, the encryption procedure becomes
\begin{equation}
\mymatrix{x}^*(t)=\mymatrix{s}(t)+\mymatrix{A}_s^{-1}\mymatrix{A}_k\mymatrix{k}(t).
\end{equation}
In this case, the encryption scheme is actually independent of the
underdetermined BSS problem. In addition, as we shown later in
Sec.~\ref{section:CODA}, the key signals can be totally circumvented
in a ciphertext-only differential attack, so the mixing matrix
$\mymatrix{A}$ must be kept as the secret key. Thus, in this paper
we assume that the secret key consists of both $\mathrm{I}_0$ and
$\mymatrix{A}$.

In \cite{Lin:BSS_IE:IEE_EL2002, Lin:BSS_IE:ICNNSP2003,
Lin:BSS_IE:CASSET2004, Lin:BSS_SIE:ISNN2005, Lin:BSS_IE:ISNN2006},
the BSS-based encryption scheme was mainly designed to encrypt $P$
images simultaneously, where $s_i(t)$ is the $t$-th pixel in the
$i$-th image. In \cite{Lin:BSS_SE:ICCCAS2004,
Lin:BSS_SE:IEEETCASI2006}, the encryption scheme was suggested to
encrypt a single speech, each frame of which is divided into $P$
segments and $s_i(t)$ is the $t$-th sample in the $i$-th segment.
This encryption scheme can also be applied for a single image, by
dividing it into $P$ blocks of the same size. To facilitate the
following discussion, we assume that the encryption scheme is used
to encrypt a single plaintext with $P$ segments of equal size.

In Sec.~VII of \cite{Lin:BSS_SE:IEEETCASI2006}, Lin et al. claimed
that the BSS-based encryption scheme is secure against most modern
cryptographical attacks, including the ciphertext-only attack, the
known-plaintext attack, and the chosen-plaintext attack. In next
section we will show that this claim is problematic.

\section{Cryptanalysis}
\label{section:Cryptanalysis}

Before introducing the cryptanalytic results, let us see how large
the key space is. In Lin et al.'s papers, each element of
$\mymatrix{A}$ is within the interval $[-1,1]$. Then, assuming that
each element in $\mymatrix{A}$ has $R$ possible values\footnote{The
value of $R$ is determined by the finite precision under which the
cryptosystem is realized. For example, if the cryptosystem is
implemented with $n$-bit fixed-point arithmetic, $R=2^n$; if it is
implemented with IEEE floating-point arithmetic, $R\approx 2^{31}$
(single-precision) or $R\approx 2^{63}$ (double-precision)
\cite{IEEEStandard754:Floating-Point}, where note that the sign bit
of the floating-point number is always negative.}, the number of all
possible mixing matrix $\mymatrix{A}$ is $R^{P(P+Q)}$. Furthermore,
assuming that the bit size of $\mathrm{I}_0$ is $L$, the size of the
whole key space is $R^{P(P+Q)}2^L$. When $Q=P$ and
$\mymatrix{A}=[\mymatrix{B},\beta\mymatrix{B}]$, the size of the
whole key space is $R^{P^2}2^L$. Later we will show that the real
size of the key space is much smaller than this estimation, due to
some essential security defects of the BSS-based encryption scheme.
We will also point out that the encryption scheme under study is not
secure against known/chosen-plaintext attack and chosen-ciphertext
attack.

\subsection{Ciphertext-Only Attack}

\subsubsection{Divide-and-Conquer (DAC) Attack}

Rewriting Eq.~(\ref{equation:decryption}) in the following form:
\begin{equation}
\mymatrix{s}(t)=\mymatrix{\hat{A}}\mymatrix{x}_k(t),
\end{equation}
where
$\mymatrix{x}_k(t)=[x_1(t),\cdots,x_P(t),k_1(t),\cdots,k_Q(t)]^T$
and
\[
\mymatrix{\hat{A}}=\mymatrix{A}_s^{-1}\left[\mymatrix{I},-\mymatrix{A}_k\right]=\left[\mymatrix{A}_s^{-1},-\mymatrix{A}_s^{-1}\mymatrix{A}_k\right].
\]
From the above equation, to recover $x_i(t)$, one only needs to know
$\mymatrix{k}(t)$ and the $i$-th row of $\mymatrix{\hat{A}}$. In
other words, when the BSS-based encryption scheme is used to encrypt
$P$ independent plaintexts, the $i$-th plaintext can be exactly
recovered with the knowledge of $\mathrm{I}_0$ and the $i$-th row of
$\mymatrix{\hat{A}}$. A similar result can be obtained when $P$
segments of one single plaintext is encrypted with the encryption
scheme. This fact means that $P$ rows of $\mymatrix{\hat{A}}$ can be
separately broken with a divide-and-conquer (DAC) attack. As a
result, the size of the key space is reduced to be $PR^{(P+Q)}2^L$.
When $Q=P$ and $\mymatrix{A}=[\mymatrix{B},\beta\mymatrix{B}]$, it
becomes $PR^{P}2^L$.

\subsubsection{Low Sensitivity to $\mymatrix{A}$}
\label{section:Low_Sensitivity_A}

From the cryptographical point of view, given two distinct keys,
even if their difference is the minimal value under the current
finite precision, the encryption and decryption results of a good
cryptosystem should still be completely different. In other words,
this cryptosystem should have a very high sensitivity to the secret
key \cite{Schneier:AppliedCryptography96}. Unfortunately, the
BSS-based encryption scheme does not satisfy this security
principle, because the involved matrix computation is not
sufficiently sensitive to matrix mismatch. Given two matrices
$\mymatrix{A}_1$ and $\mymatrix{A}_2$ of size $M\times N$, if the
maximal difference of all elements is $\varepsilon$, then one can
easily deduce that each element of
$|\mymatrix{A}_1\mymatrix{s}(t)-\mymatrix{A}_2\mymatrix{s}(t)|$ is
not greater than $N\max(\mymatrix{s}(t))\varepsilon$. As a result,
the matrix $\mymatrix{A}$ can be approximately guessed under a
relatively large finite precision $\varepsilon$, still maintaining
an acceptable quality of the recovered plaintexts. This immediately
leads to a significant reduction of the size of the key space: from
$PR^{(P+Q)}2^L$ to $P\lceil 2/\varepsilon\rceil^{(P+Q)}2^L$, where
$\lceil 2/\varepsilon\rceil^{(P+Q)}\ll R^{(P+Q)}$.

\begin{figure}[!htbp]
\center
\begin{minipage}{\figwidth}
\center
\includegraphics[width=\textwidth]{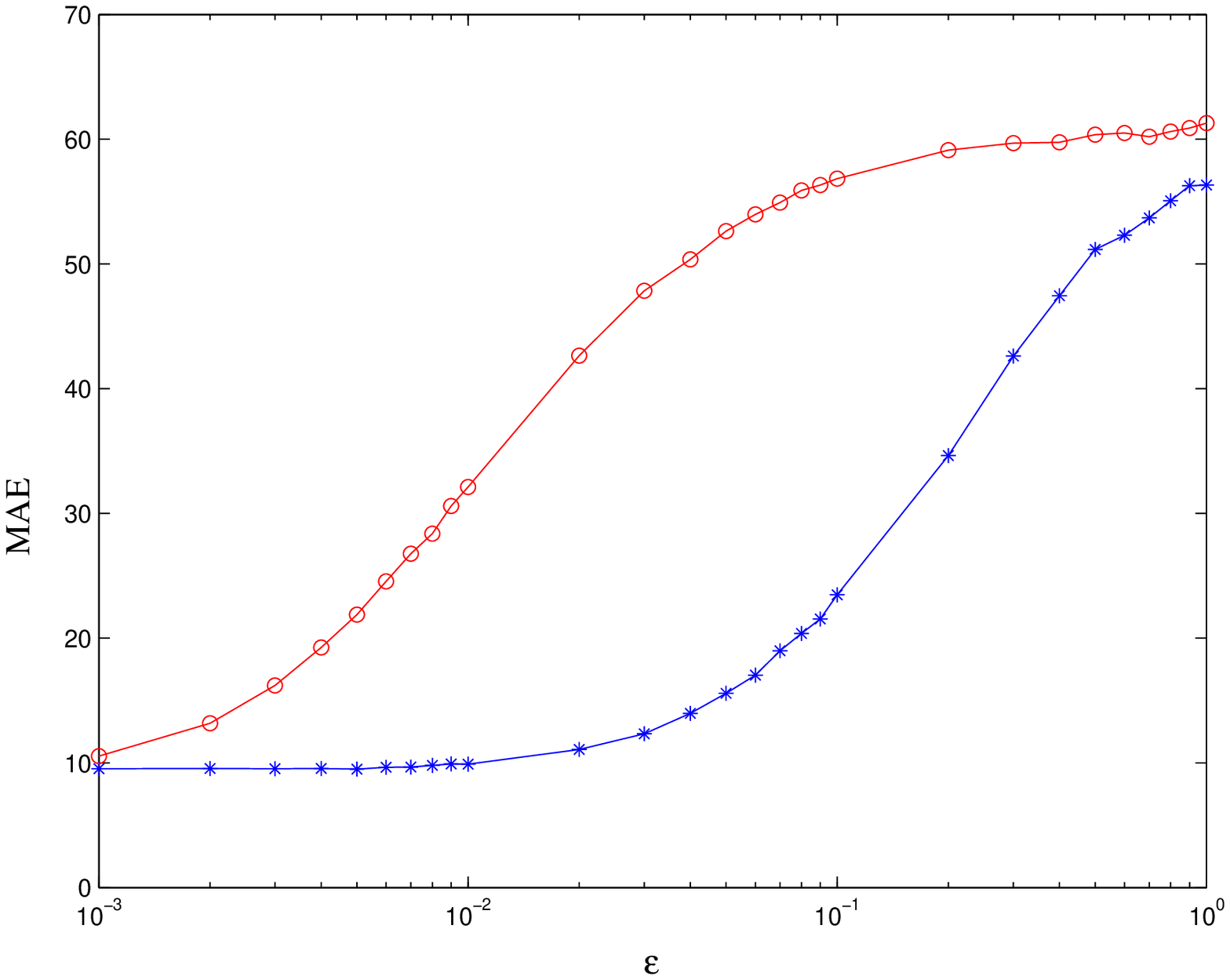}
Legend: $\textcolor{blue}{\ast}$ -- $P=Q=4$;
$\textcolor{red}{\circ}$ -- $P=4$ and
$\mymatrix{A}=[\mymatrix{B},\beta\mymatrix{B}]$ ($\beta=10$).\\
a)
\end{minipage}
\begin{minipage}{\figwidth}
\center
\includegraphics[width=\textwidth]{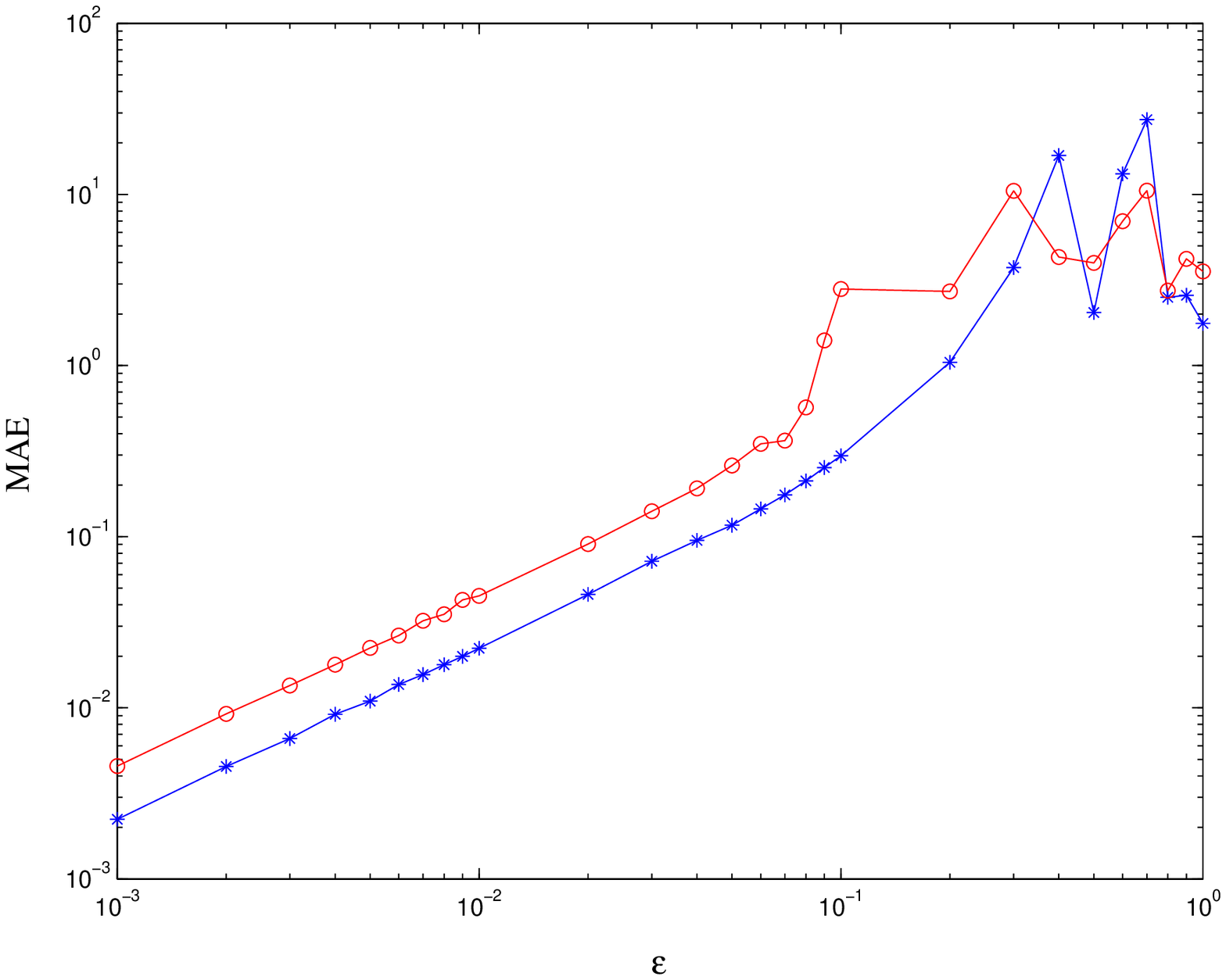}
Legend: $\textcolor{blue}{\ast}$ -- $P=Q=4$;
$\textcolor{red}{\circ}$ -- $P=4$ and
$\mymatrix{A}=[\mymatrix{B},\beta\mymatrix{B}]$ ($\beta=2$).\\
b)
\end{minipage}
\caption{The experimental relationship between the recovery error
and the value of $\varepsilon$: a) the plaintext is a digital image
``Lenna" (Fig.~\ref{figure:HVS_error}a); b) the plaintext is a
speech file ``one.wav" that corresponds to the pronunciation of the
English word ``one" (from Merriam-Webster Online Dictionary,
http://www.m-w.com).}\label{figure:epsilon_MAE}
\end{figure}

The above low sensitivity can be easily verified with experiments
described as follows:
\begin{itemize}
\item
\textit{Step 1}: for a randomly-generated key
$(\mymatrix{A},\mathrm{I}_0)$, calculate the ciphertext
$\mymatrix{x}(t)$ corresponding to a plaintext $\mymatrix{s}(t)$;

\item
\textit{Step 2}: with another mismatched key
$(\mymatrix{A}+\varepsilon\mymatrix{R},\mathrm{I}_0)$, decrypt
$\mymatrix{x}(t)$ to get $\mymatrix{\tilde{s}}(t)$ -- an estimated
version of $\mymatrix{s}(t)$, where $\varepsilon\in(0,1)$ and
$\mymatrix{R}$ is a $P\times(P+Q)$ random $(1,-1)$-matrix.
\end{itemize}
For each value of $\varepsilon$, the second step was repeated for
100 times to get a mean value of the recovery error (measured in MAE
-- mean absolute error)\footnote{When the plaintext is a digital
image with 256 gray scales, we first calibrate each sub-image into
the range $\{0,\cdots,255\}$ and then calculate the recovery error
of the whole image.}. Then, we can observe the relationship between
the recovery error and the value of $\varepsilon$.
Figure~\ref{figure:epsilon_MAE} shows the experimental results when
the plaintexts are a digital image and a speech file, respectively.

\begin{figure}[!htbp]
\center
\includegraphics[width=\figwidth]{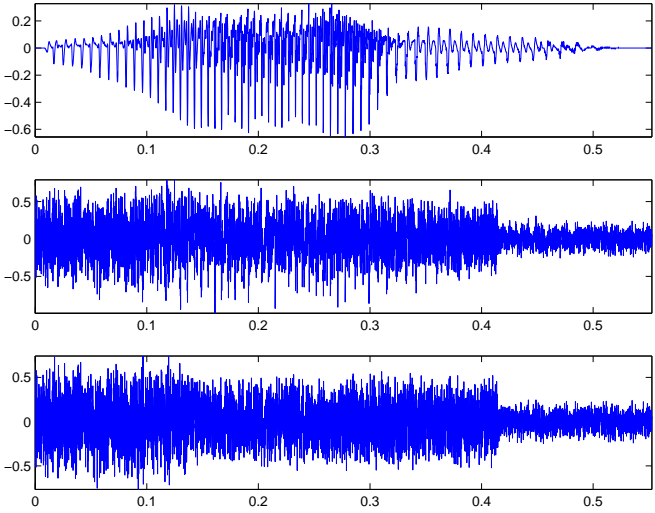}
\caption{An example of human capability against large noises in
speech. From top to bottom: the original plain-speech ``one.wav",
the recovered speech, the recovery error (MAE=0.164103). For
reader's sake, the recovered speech is posted online at
http://www.hooklee.com/Papers/Data/BSSE/one\_MAE=0.164103.wav.}\label{figure:HAS_error}
\end{figure}

\begin{figure}[!htbp]
\center
\begin{minipage}{\imagewidth}
\center
\includegraphics[width=\imagewidth]{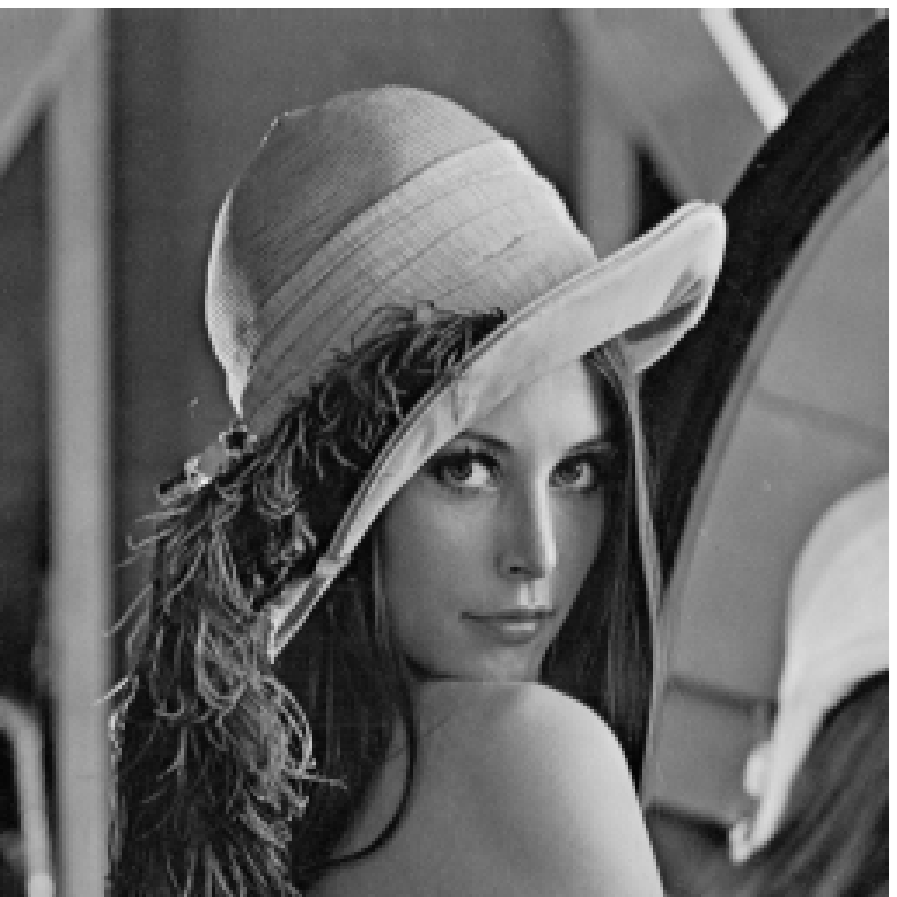}
a)
\end{minipage}
\begin{minipage}{\imagewidth}
\center
\includegraphics[width=\imagewidth]{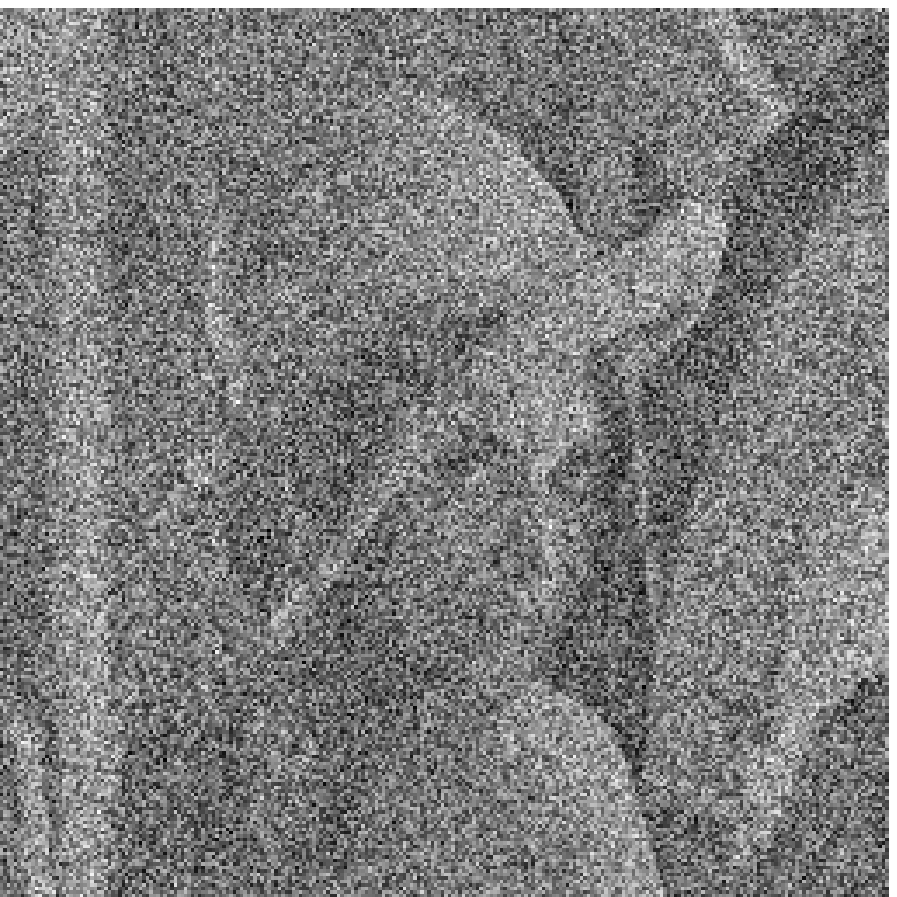}
b)
\end{minipage}
\caption{An example of human capability against large noises in
images: a) the original plain-image ``Lenna"; b) the recovered image
(MAE=47.6913).}\label{figure:HVS_error}
\end{figure}

The experimental results confirms that a mismatched key can
approximately recover the plaintext. Considering that humans have a
good capability of resisting errors in images and speech, even
relatively large errors may not be able to prevent a human attacker
from recognizing the plain-image or plain-speech. Thus, the value of
$\varepsilon$ may be relatively large. When $P=4$,
$\mymatrix{A}=[\mymatrix{B},\beta\mymatrix{B}]$ and
$\varepsilon=0.1$, we give two examples of such recognizable
plaintexts with relatively large errors in
Figs.~\ref{figure:HAS_error} and \ref{figure:HVS_error}.

From the above experimental results, we can exhaustively search for
an approximate version of $\mymatrix{A}$ under the finite precision
$\varepsilon=0.01\sim 0.1$. Such an approximate version of
$\mymatrix{A}$ is then used to roughly reveal the plaintext.
Considering the searching complexity is
$O\left(\varepsilon^{-(P+Q)}\right)$, such an exhaustive search is
feasible when $P,Q$ is not very large\footnote{In
\cite{Lin:BSS_IE:IEE_EL2002, Lin:BSS_IE:ICNNSP2003,
Lin:BSS_IE:CASSET2004, Lin:BSS_SIE:ISNN2005, Lin:BSS_IE:ISNN2006,
Lin:BSS_SE:ICCCAS2004, Lin:BSS_SE:IEEETCASI2006}, small values are
used in all examples: $P=2$ or 4 and $Q\leq P$.}. When $P=2$ and
$\mymatrix{A}=[\mymatrix{B},\beta\mymatrix{B}]$, we carried out a
large number of experiments in the following steps:
\begin{itemize}
\item
\textit{Step 1}: for a randomly-generated key
$(\mymatrix{B},\mathrm{I}_0)$, calculate the ciphertext
$\mymatrix{x}(t)$ corresponding to a plaintext $\mymatrix{s}(t)$;

\item
\textit{Step 2}: randomly generate a matrix $\mymatrix{R}$ (each
element over the interval $[-1,1]$), and then decrypt
$\mymatrix{x}(t)$ with the guessed key $(\mymatrix{R},\mathrm{I}_0)$
to get $\mymatrix{\tilde{s}}(t)$;

\item
\textit{Step 3}: repeat \textit{Step 2} for $r$ rounds, output the
recovered plaintext $\mymatrix{\tilde{s}}^*(t)$, every segment of
which corresponds to the best recovery performance in all the $r$
rounds;

\item
\textit{Step 4}: for the $i$-th segment of
$\mymatrix{\tilde{s}}^*(t)$, find the corresponding matrix
$\mymatrix{R}$, extract its $i$-th row of its inverse
$\mymatrix{R}^{-1}$ to form the $i$-th row of
$\mymatrix{\tilde{B}}^{-1}$, the inverse of an estimation of the
original matrix $\mymatrix{B}$.
\end{itemize}
Assuming that the target finite precision is $\varepsilon>0$, the
interval $[-1,1]$ is divided into $n_{\varepsilon}=\lceil
2/\varepsilon\rceil$ sub-intervals. Without loss of generality,
assuming that $2/\varepsilon$ is an integer, then each sub-interval
is of equal size. Thus, if the element in the random matrix
$\mymatrix{R}$ has a uniform distribution over $[-1,1]$, the
probability that $|r_{i,j}-a_{i,j}|<\varepsilon$ occurs at least one
time in $r$ rounds of experiment is
$p(n_{\varepsilon},r)=1-(1-1/n_{\varepsilon})^r$, where $r_{i,j}$
and $a_{i,j}$ are the $(i,j)$-th elements of $\mymatrix{R}$ and
$\mymatrix{A}$, respectively. One can easily deduce that
$p(n_{\varepsilon},r)$ is an increasing function with respect to $r$
and
\begin{eqnarray*}
p(n_{\varepsilon},n_{\varepsilon})>
\lim_{n_{\varepsilon}\to\infty}p(n_{\varepsilon},n_{\varepsilon}) &
= &
1-\lim_{n_{\varepsilon}\to\infty}(1-1/n_{\varepsilon})^{n_{\varepsilon}}\\
& = & 1-e^{-1}\approx 0.6321,
\end{eqnarray*}
which leads to the result that $p(n_{\varepsilon},r)>1-e^{-1}$ when
$r\geq n_{\varepsilon}$. In other words, with $r\geq
n_{\varepsilon}$ experiments, it is a high-probability event that we
have at least one $r_{i,j}$ ``equal" to $a_{i,j}$ under the finite
precision $\varepsilon$. To get an approximate estimation of the
$i$-th row of $\mymatrix{A}$, we can see that
$r=O\left(n_\varepsilon^P\right)$ rounds of experiment are needed.

\begin{figure}[!htbp]
\center
\includegraphics[width=\figwidth]{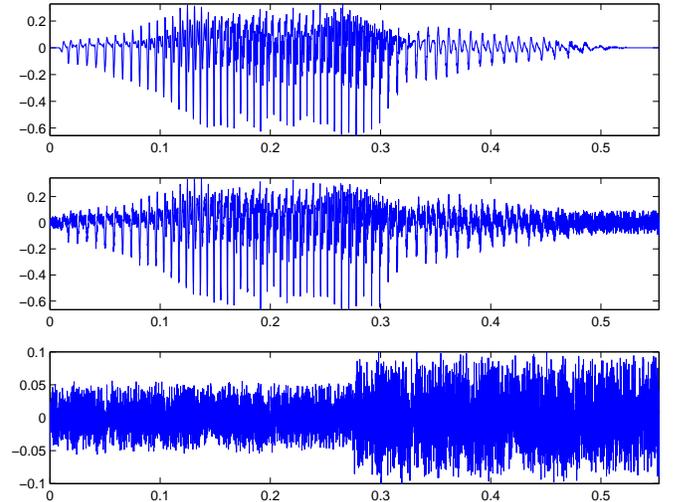}
\caption{A recovered speech in one 50,000-round experiment of
exhaustively guessing $\mymatrix{A}$ when $P=2$ and
$\mymatrix{A}=[\mymatrix{B},\beta\mymatrix{B}]$. From top to bottom:
the original plain-speech ``one.wav", the recovered speech (MANE of
each segment: 0.0469, 0.0521), the recovery error. For reader's
sake, the recovered speech is posted online at
http://www.hooklee.com/Papers/Data/BSSE/one\_MANE=0.0469-0.0521.wav.}\label{figure:B_search_B_one}
\end{figure}

\begin{figure}[!htbp]
\center
\begin{minipage}{\imagewidth}
\center
\includegraphics[width=\imagewidth]{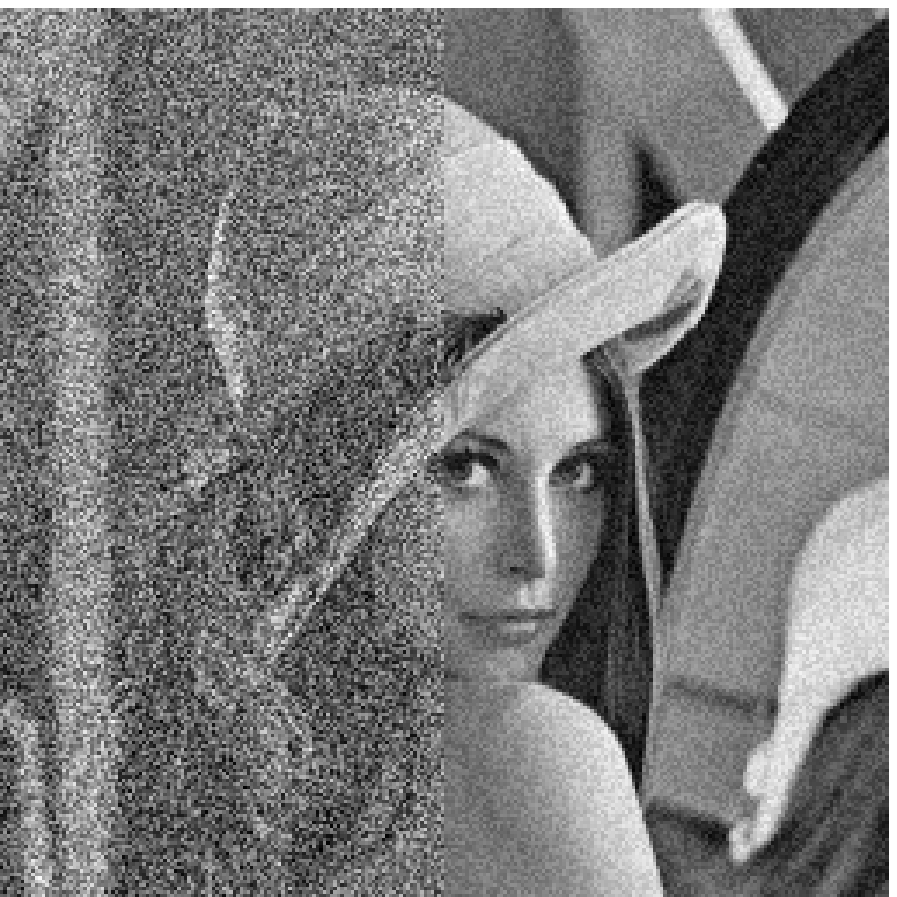}
a)
\end{minipage}
\begin{minipage}{\imagewidth}
\center
\includegraphics[width=\imagewidth]{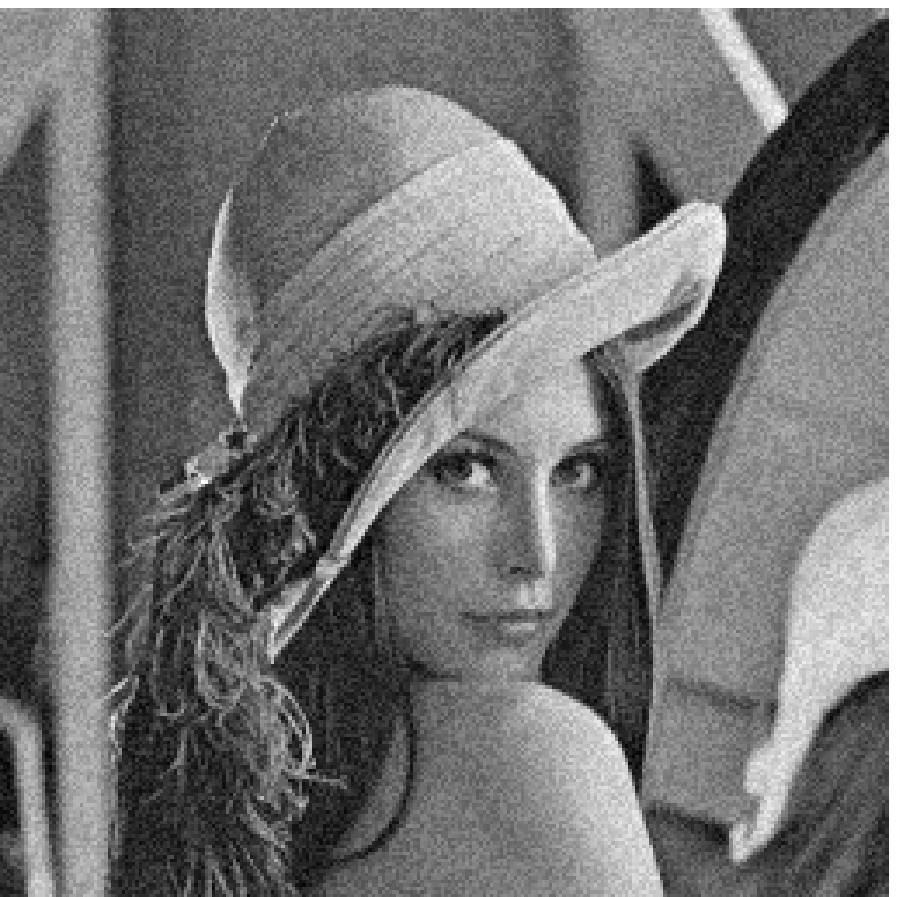}
b)
\end{minipage}
\caption{Two recovered plain-images in our experiments of
exhaustively guessing $\mymatrix{B}$ when $P=2$ and
$\mymatrix{A}=[\mymatrix{B},\beta\mymatrix{B}]$: a) $r=1,000$ (MANE
of each segment: 39.7491, 14.9373); b) $r=10,000$ (MANE of each
segment: 16.3888, 15.1722).}\label{figure:B_search_B_Lenna}
\end{figure}

Apparently, the above steps actually simulate the process of a real
ciphertext-only attack that tries to reveal the plaintext and to
exhaustively guess $\mymatrix{B}^{-1}$ (under the assumption that
$\mathrm{I}_0$ has been known). Note that MAE cannot be calculated
to evaluate the recovery performance in a real attack, in which one
does not know the plaintext. Fortunately, exploiting the large
information redundancy existing in natural images and speech, one
can turn to use some other measures to reflect the recovery
performance of each segment of $\mymatrix{\tilde{s}}(t)$. In our
experiments, we use a measure called MANE (mean absolute neighboring
error), which is defined as follows for the $i$-th segment of
$\mymatrix{\tilde{s}}(t)$
\begin{equation}
\frac{1}{T-2}\sum_{t=2}^{T-1}\frac{|\tilde{s}_i(t)-\tilde{s}_i(t-1)|+|\tilde{s}_i(t)-\tilde{s}_i(t+1)|}{2},
\end{equation}
where $T$ denotes the segment length. In
Figs.~\ref{figure:B_search_B_one} and \ref{figure:B_search_B_Lenna},
one recovered plain-speech and two recovered plain-images are shown
for demonstration. One can see that $r=O(10,000)$ (or
$\varepsilon\approx 0.01$) is sufficient to get a good estimation of
the plaintext.

Note that for 2-D images the above 1-D MANE may be generalized to
include more neighboring pixels, thus achieving a more accurate
description of the recovery performance. In addition, multiple
quality factors can be employed to further increase the efficiency
of evaluation of the recovery performance.

\subsubsection{Low Sensitivity to $\mymatrix{k}(t)$}

Due to the same reason of the low sensitivity to $\mymatrix{A}$, one
can deduce that the BSS-based encryption scheme is also insensitive
to the key signal $\mymatrix{k}(t)$. Given two key signals
$\mymatrix{k}_1(t)$ and $\mymatrix{k}_2(t)$, if the maximal
difference of all elements is $\varepsilon$, each element of
$|\mymatrix{A}_k\mymatrix{k}_1(t)-\mymatrix{A}_k\mymatrix{k}_2(t)|$
is not greater than
$Q\max(|\mymatrix{A}_k|)\varepsilon=Q\varepsilon$. Since
$\mymatrix{k}(t)$ itself is not part of the secret key, but
generated from $\mathrm{I}_0$, this problem does not have much
negative influence on the security of the whole cryptosystem against
ciphertext-only attacks.

\subsubsection{Low Sensitivity to Plaintext}
\label{section:Low_Sensitivity_Plaintext}

Another cryptographical property required by a good cryptosystem is
that the encryption is very sensitive to plaintext, i.e., the
ciphertexts of two plaintexts with a slight difference should be
much different \cite{Schneier:AppliedCryptography96}. However, this
property does not hold for the BSS-based encryption scheme. Given
two key signals $\mymatrix{s}_1(t)$ and $\mymatrix{s}_2(t)$, if the
maximal difference of all elements is $\varepsilon$, each element of
$|\mymatrix{A}_s\mymatrix{s}_1(t)-\mymatrix{A}_s\mymatrix{s}_2(t)|$
is not greater than
$P\max(|\mymatrix{A}_s|)\varepsilon=P\varepsilon$. When the same
secret key is used to encrypt two close-correlated plaintexts, such
as a plaintext and its watermarked version, this security defect
means that the exposure of one plaintext leads to the revealment of
both.

\subsubsection{Differential Attack}
\label{section:CODA}

\begin{figure}[!htbp]
\center
\includegraphics[width=\figwidth]{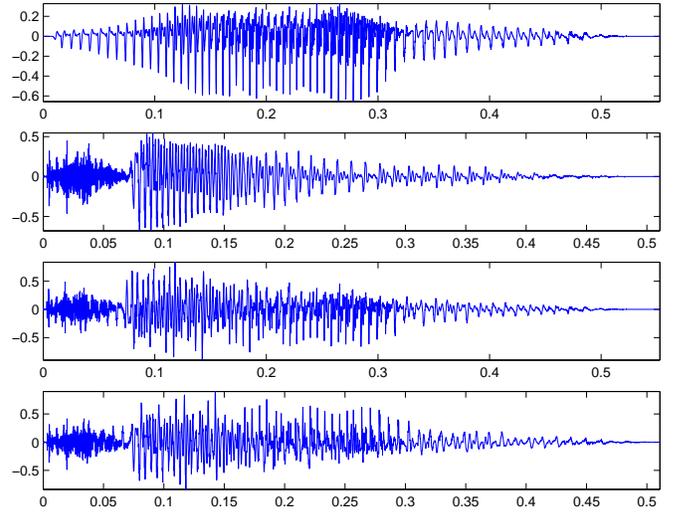}
\caption{Differentials of two plain-speech files. From top to
bottom: the first speech ``one.wav", the second speech ``two.wav",
the differential one-two, the differential two-one. For readers'
sake, the two differential speech files are posted online at
http://www.hooklee.com/Papers/Data/BSSE/one-two.wav and
http://www.hooklee.com/Papers/Data/BSSE/two-one.wav.}\label{figure:Differential_Speech}
\end{figure}

\begin{figure}[!htbp]
\center
\begin{minipage}{\imagewidth}
\center
\includegraphics[width=\imagewidth]{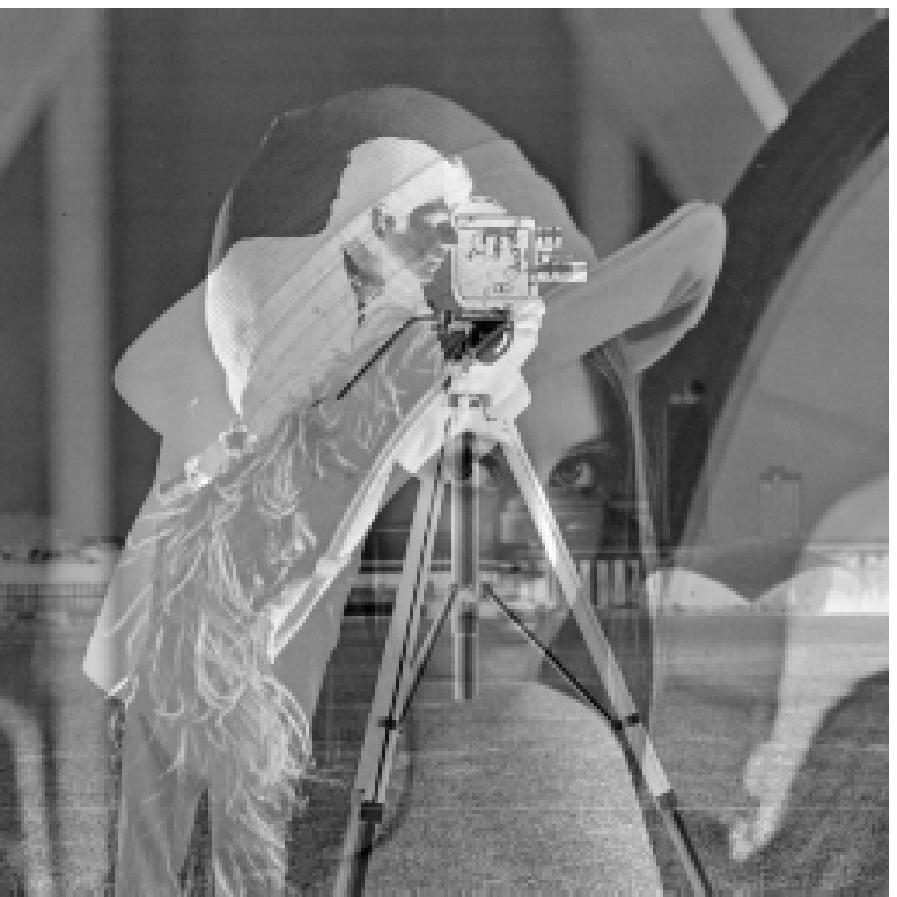}
a)
\end{minipage}
\begin{minipage}{\imagewidth}
\center
\includegraphics[width=\imagewidth]{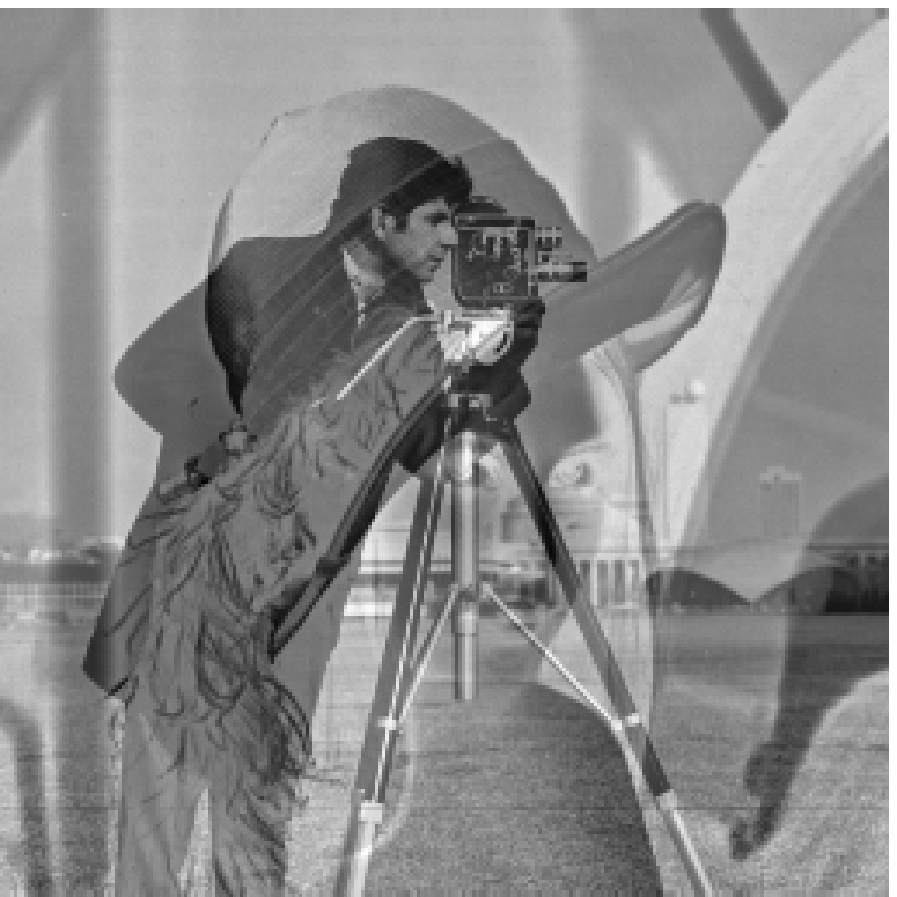}
b)
\end{minipage}
\caption{Differentials of two plain-images, ``Lenna" and
``cameraman": a) Lenna-cameraman; b)
cameraman-Lenna.}\label{figure:Differential_Images}
\end{figure}

Given two plaintexts $\mymatrix{s}^{(1)}(t)$ and
$\mymatrix{s}^{(2)}(t)$, if they are encrypted with the same key
$(\mymatrix{A},\mathrm{I}_0)$, we can get the following formula from
Eq.~(\ref{equation:encryption}):
\begin{equation}\label{equation:Delta_xt}
\Delta_{\mymatrix{x}}(t)=\mymatrix{A}_s\Delta_{\mymatrix{s}}(t),
\end{equation}
where
$\Delta_{\mymatrix{x}}(t)=\mymatrix{x}^{(1)}(t)-\mymatrix{x}^{(2)}(t)$
and
$\Delta_{\mymatrix{s}}(t)=\mymatrix{s}^{(1)}(t)-\mymatrix{s}^{(2)}(t)$.
Note that $\mymatrix{A}_k\mymatrix{k}(t)$ disappears in the above
equation. This means that from the differential viewpoint only
$\mymatrix{A}_s$ is the secret key, i.e., $\mathrm{I}_0$ is removed
from the key. Considering the low sensitivity of the encryption
scheme to $\mymatrix{A}$, under finite precision $\varepsilon$ the
key space becomes $O\left(P\varepsilon^{-P}\right)$, and one might
exhaustively search $\mymatrix{A}_s$ to recover the plaintext
differential as follows:
\begin{equation}\label{equation:Delta_st}
\Delta_{\mymatrix{s}}(t)=\mymatrix{A}_s^{-1}\Delta_{\mymatrix{x}}(t).
\end{equation}
From the obtained plaintext differential, one can get a mixed view
of the two interested plaintexts, from which both plaintexts may be
completely recognizable by humans. See
Figs.~\ref{figure:Differential_Speech} and
\ref{figure:Differential_Images} for four plaintext differentials of
two speech files and two images.

\begin{figure}[!htbp]
\center
\includegraphics[width=\imagewidth]{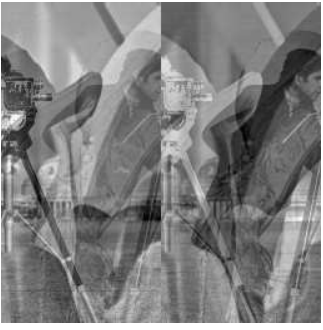}
\caption{One obtained plain-image differential when $\mymatrix{A}_s$
and $\mymatrix{\tilde{A}}_s$ have a relatively large mismatch as
shown in
Eq.~(\ref{equation:As_As2}).}\label{figure:image-large-mismatch}
\end{figure}

\begin{figure}[!htbp]
\center
\includegraphics[width=\figwidth]{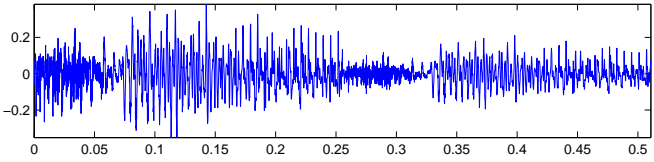}
\caption{One obtained plain-speech differential when
$\mymatrix{A}_s$ and $\mymatrix{\tilde{A}}_s$ have a relatively
large mismatch. For readers' sake, this differential speech is
posted online at
http://www.hooklee.com/Papers/Data/BSSE/two-one-large-mismatch.wav.}\label{figure:speech-large-mismatch}
\end{figure}

Denoting the guessed matrix by $\mymatrix{\tilde{A}}_s$, we have
\begin{equation}
\tilde{\Delta}_{\mymatrix{s}}(t)=\mymatrix{\tilde{A}}_s^{-1}\Delta_{\mymatrix{x}}(t)=
\mymatrix{\tilde{A}}_s^{-1}\mymatrix{A}_s\Delta_{\mymatrix{s}}(t).
\end{equation}
Apparently, if $\mymatrix{\tilde{A}}_s\neq\mymatrix{A}_s$, the
obtained plaintext differential $\tilde{\Delta}_{\mymatrix{s}}(t)$
will have an inter-segment mixture, which may make the recognition
of the two plaintexts more difficult. Fortunately, when $P$ is
relatively small, such an inter-segment mixture may not be too
severe to prevent the recognition of the two plaintexts by humans.
More importantly, our experiments showed that humans can even be
able to recognize the two plaintexts even when the mismatch between
$\mymatrix{\tilde{A}}_s$ and $\mymatrix{A}_s$ is not very small.
When $P=2$,
\begin{equation}
\mymatrix{A}_s=\left[
\begin{matrix}
0.7123 & -0.4272\\
0.1958 & 0.1295
\end{matrix}\right]\mbox{, }
\mymatrix{\tilde{A}}_s=\left[
\begin{matrix}
0.5914 & 0.9527\\
0.5726 & 0.1437
\end{matrix}\right],\label{equation:As_As2}
\end{equation}
a plaintext differential obtained in our experiments is shown in
Fig.~\ref{figure:image-large-mismatch}. One can see that both
plain-images, ``Lenna" and ``cameraman", can still be roughly
recognized from such a heavily mixed differential. Another obtained
plain-speech differential for ``one.wav" and ``two.wav", is shown in
Fig.~\ref{figure:speech-large-mismatch}, from which the two English
words (``one" and ``two") are also perceptible.

\begin{figure}[!htbp]
\center
\begin{minipage}{\imagewidth}
\center
\includegraphics[width=\imagewidth]{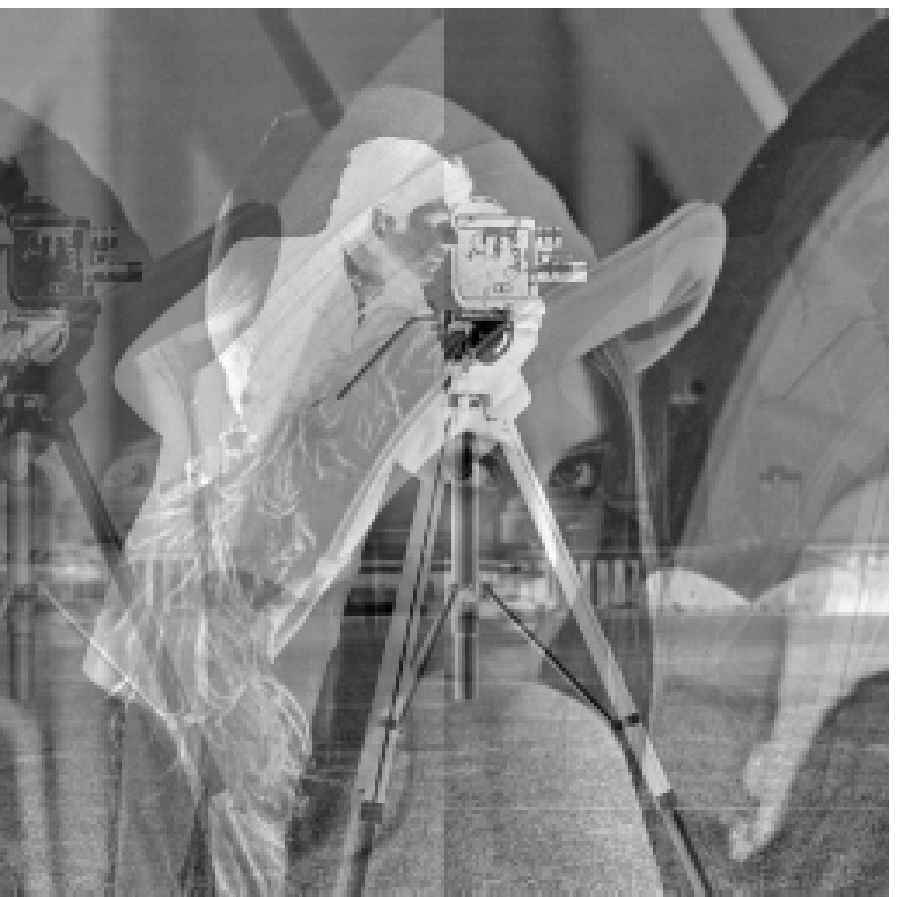}
a)
\end{minipage}
\begin{minipage}{\imagewidth}
\center
\includegraphics[width=\imagewidth]{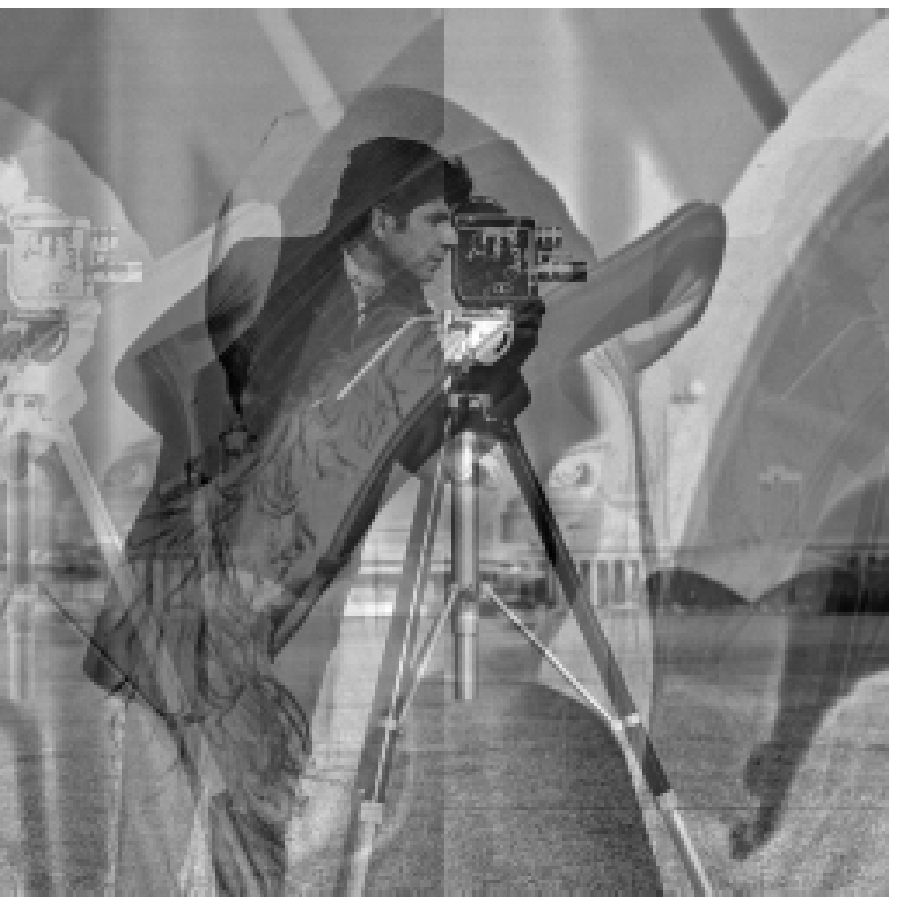}
b)
\end{minipage}
\caption{A visually-optimal result obtained in 100 plain-image
differentials: a) the differential; b) the negative image of the
differential.}\label{figure:1_in_100}
\end{figure}

In this differential attack, the quality evaluation factors (such as
MANE) used in Sec.~\ref{section:Low_Sensitivity_A} is not suitable
to automatically determine the best result in many plaintext
differentials, because each segment of the obtained plaintext
differential is also a natural signal with abundant information
redundancy. Instead, one has to output all obtained differentials,
and check them with naked eyes or ears to find a
perceptually-optimal result with the least inter-segment mixture.
Figure~\ref{figure:1_in_100} shows such a result in 100 plain-image
differentials when $P=2$ and $\mymatrix{A}$ follows
Eq.~(\ref{equation:As_As2}). By checking each segment separately and
combine the $P$ optimal segments together, one can further get a
better result with less inter-segment mixture.

While this differential attack works well for $P=2$ as shown above,
it will become infeasible when $P$ is sufficiently large, due to the
following facts: 1) the inter-segment mixture is too severe; 2) the
complexity of checking all $O\left(\varepsilon^{-P}\right)$
differentials is beyond humans' capability.

\subsection{Known-Plaintext Attack}

In this kind of attack, one can access to a number of plaintexts
that are encrypted with the same key. Then, from
Eq.~(\ref{equation:Delta_xt}), with $P$ plaintext differentials, one
immediately knows that the mixing matrix can be uniquely determined
as follows:
\begin{equation}
\mymatrix{A}_s=\Delta_{\mymatrix{X}}(t)(\Delta_{\mymatrix{S}}(t))^{-1},
\end{equation}
where $\Delta_{\mymatrix{S}}(t)$ and $\Delta_{\mymatrix{X}}(t)$ are
$P\times P$ matrices, constructed row by row from the $P$ plaintext
differentials and the corresponding ciphertext differentials,
respectively. Then, $\mymatrix{A}_k\mymatrix{k}(t)$ can be further
solved from any plaintext and its ciphertext:
\begin{equation}
\mymatrix{A}_k\mymatrix{k}(t)=\mymatrix{x}(t)-\mymatrix{A}_s\mymatrix{s}(t).
\end{equation}
Now, $(\mymatrix{A}_s,\mymatrix{A}_k\mymatrix{k}(t))$ can be used to
recover other plaintexts encrypted by the same key
$(\mymatrix{A},\mathrm{I}_0)$. Note that
$\mymatrix{A}_k\mymatrix{k}(t)$ has a finite length determined by
the maximal length of all known plaintexts, so
$(\mymatrix{A}_s,\mymatrix{A}_k\mymatrix{k}(t))$ can only recover
plaintexts under this finite length.

When $\mymatrix{A}=[\mymatrix{B},\beta\mymatrix{B}]$, the key
signals can also be determined:
\begin{equation}
\mymatrix{k}(t)=\frac{\mymatrix{s}(t)-\mymatrix{B}^{-1}\mymatrix{x}(t)}{\beta}.
\end{equation}
If the PRNG used is not cryptographically strong (such as
LFSR\cite{Schneier:AppliedCryptography96}), it may be possible to
further derive the secret seed $\mathrm{I}_0$, thus completely
breaking the BSS-based encryption scheme.

Note that $n$ distinct plaintexts can generate
$\binom{n}{2}=n(n-1)/2$ plaintext differentials. Solving the
inequality $n(n-1)/2\geq P$, one can get the number of required
plaintexts to yield at least $P$ plaintext differentials:
\begin{equation}
n\geq\left\lceil\sqrt{P-1/4}+1/2\right\rceil\approx\sqrt{P}.
\end{equation}

\subsection{Chosen-Plaintext/Ciphertext Attack}

In chosen-plaintext attack, one can freely choose a number of
plaintexts and observe the corresponding ciphertexts, while in
chosen-ciphertext attack, one can freely choose a number of
ciphertexts and observe the corresponding plaintexts. So in these
attacks, one can choose $P$ plaintext differentials easily, which
means that the above differential known-plaintext attack still works
in the same way.

\section{Discussion}
\label{section:Discussion}

As we pointed out in last section, the BSS-based encryption scheme
is always insecure against plaintext attack. So the secret key
cannot be repeatedly used in any case. This means that the
encryption scheme has to work like a common stream cipher, by
changing the secret key for each distinct plaintext. However, in
this case, $\mymatrix{k}(t)$ (equivalently, the secret seed
$\mathrm{I}_0$) is enough to provide a high level of security, since
$\mymatrix{k}(t)$ satisfies the cryptographical properties in a
perfectly secure one-time-a-pad cipher (see Sec.~V.B of
\cite{Lin:BSS_SE:IEEETCASI2006}). Then, the mixing matrix
$\mymatrix{A}$ becomes excessive.

Even when one wants to add a second defense to potential attacks by
applying the BSS mixing, the low sensitivity of
encryption/decryption to the mixing matrix $\mymatrix{A}$ (recall
Sec.~\ref{section:Low_Sensitivity_A}) makes this goal less useful.
As a result, with the current encryption design, the BSS model does
not play a key role in the security of the scheme. The real core of
the encryption scheme is the embedded PRNG that is in charge of
generating the key signals masking the plaintexts.

If one wants to use the BSS-based encryption scheme with repeatedly
used key, some essential modifications have to be made to reinforce
the security against various attacks. Following the cryptanalytic
results given in last section, we suggest adopting two
coutermeasures simultaneously: 1) use a sufficiently large $P$; 2)
like the design of most modern block ciphers
\cite{Schneier:AppliedCryptography96}, iterate the BSS-based
encryption for many rounds to avoid the original scheme's low
sensitivity to the secret key and plaintext. It is obvious that both
countermeasures will significantly influence the
encryption/decryption speed of the encryption scheme. It seems
doubtful if such an enhanced encryption scheme will have any
advantages compared with other multiple-round block ciphers,
especially AES \cite{NIST:AES2001} that can be optimized to run with
a very high rate on PCs \cite{Gladman:AESCode}.

Finally, it deserve mentioning that the original BSS-based
encryption scheme can be used to realize \textbf{lossy} decryption,
an interesting feature that may find useful in some real
applications\footnote{Another scheme is a matrix-based image
scrambling system proposed in \cite{Ville:ISWBE:CASVT2004}, as
pointed out in \cite{ShujunLi:AttackISWBE2006}.}. This feature means
that an encryption scheme can still (maybe roughly) recover the
plaintext even when there are some errors in the ciphertexts. An
typical use of this feature is that the ciphertext can be compressed
with some lossy algorithms to save the required storage in local
computers or the channel width for transmission. For the BSS-based
encryption scheme, the lossy decryption feature is ensured by low
sensitivity of decryption to ciphertext, which is due to the same
reason of the low sensitivity of encryption to plaintext (recall
Sec.~\ref{section:Low_Sensitivity_Plaintext}). However, keep in mind
that the lossy decryption feature is induced by the low sensitivity
to plaintext/ciphertext, so there is a tradeoff between this feature
and security.

\section{Conclusion}

This paper analyzes the security of an image/speech encryption
scheme based on BSS mixing technology \cite{Lin:BSS_IE:IEE_EL2002,
Lin:BSS_IE:ICNNSP2003, Lin:BSS_IE:CASSET2004, Lin:BSS_SIE:ISNN2005,
Lin:BSS_IE:ISNN2006, Lin:BSS_SE:ICCCAS2004,
Lin:BSS_SE:IEEETCASI2006}. It has been shown that this BSS-based
encryption scheme suffers from some security defects, including its
vulnerability to a ciphertext-only differential attack,
known/chosen-plaintext attack and chosen-ciphertext attack. It
remains an open problem how to apply BSS technology to construct
cryptographically strong ciphers.

\bibliographystyle{IEEEtran}
\bibliography{IEEEabrv,mypaper}

\end{document}